\documentclass[aps,prx,preprint,superscriptaddress,longbibliography]{revtex4-2}
\usepackage{multirow}
\usepackage{array}
\usepackage{soul}

\usepackage{float}
\usepackage{graphicx}
\usepackage{afterpage}
\usepackage{caption}
\usepackage{subcaption}
\usepackage[font=small,labelsep=quad,format=plain]{caption}

\usepackage[colorlinks=true,citecolor=blue,linkcolor=blue,urlcolor=blue]{hyperref}
\usepackage{bm}
\usepackage{physics}

\begin{document}

\title{Realizing symmetry-protected topological phases in a spin-1/2 chain with next-nearest neighbor hopping on superconducting qubits}

\author{Adrian~T.~K.~Tan}
\affiliation{Division of Engineering and Applied Science, California Institute of Technology, Pasadena, CA 91125, USA}
\author{Shi-Ning Sun}
\affiliation{Division of Engineering and Applied Science, California Institute of Technology, Pasadena, CA 91125, USA}
\author{Ruslan N. Tazhigulov}
\affiliation{Division of Chemistry and Chemical Engineering, California Institute of Technology, Pasadena, CA 91125, USA}
\author{Garnet Kin-Lic Chan}
\affiliation{Division of Chemistry and Chemical Engineering, California Institute of Technology, Pasadena, CA 91125, USA}
\author{Austin J. Minnich}
\email{aminnich@caltech.edu}
\affiliation{Division of Engineering and Applied Science, California Institute of Technology, Pasadena, CA 91125, USA}
\date{\today}

\begin{abstract}
The realization of novel phases of matter on quantum simulators is a topic of intense interest.  Digital quantum computers offer a route to prepare topological phases with interactions that do not naturally arise in analog quantum simulators. Here, we report the realization of symmetry-protected topological (SPT) phases of a spin-{1/2} Hamiltonian with next-nearest-neighbor hopping on up to  11 qubits on a programmable superconducting quantum processor. We  observe clear signatures of the two distinct SPT phases, such as  excitations localized to specific edges and finite string order parameters. Our work advances ongoing efforts to realize novel states of matter with exotic interactions on digital near-term quantum computers.
\end{abstract}

\begin{minipage}[h]{\textwidth}
    \maketitle
\end{minipage}
\vfill
\small{*aminnich@caltech.edu}
\newpage
\section{Introduction}

Quantum computers have long been of interest for their potential to simulate quantum many-body systems \cite{Feynnman_1982, Lloyd_1996, georgescu_2014, Altman_2021, sheng2021quantum, sawaya2021}. A recent emphasis has been  using quantum computers to treat classically challenging chemistry and condensed matter problems  \cite{McArdle_2020, Bauer_2020, cerezo_variational_2021, sheng2021quantum, sawaya2021}. Advances in near-term quantum hardware now make prototype versions of these simulations possible, for instance in the computation of the ground state properties of chemical \cite{peruzzo_2014, omalley_2016, kandala_2017, Colless_2018, mcardle_variational_2019, google_hartree-fock_2020, huggins2021unbiasing, seetharam2021} and solid-state \cite{kandala_2017,  Mazzola_2019, ma_2020, neill_2021} quantum systems as well as simulation of their real-time dynamics for closed \cite{barends_digital_2015, Lamm_2018, chiesa_2019, smith_2019, Endo2020, cirstoiu_variational_2020, Francis_2020, Chen_2020, Arute_2020, Fauseweh_2021, gibbs2021} and open \cite{Hu_2020, Lorenzo2020, hu2021, kade_2021, kamakari2021, schlimgen2021, rost2021} systems. Several recent studies have also reported the simulation of finite-temperature physics on near-term devices \cite{motta_2019, Cohn_2020, Zhu_2020, Sun_2021}.

The realization of topological phases of matter is another area of considerable interest. These phases do not fit within the Landau paradigm of local order parameters associated with symmetry breaking, and the study of their ground-state properties and excitations is an active area of research in  condensed matter physics \cite{Hasan_2010, Qi_2011, Senthil2014, Xiaogang2017}. Analog quantum simulators are able to realize some of these phases and associated phenomena such as models with topological band structures \cite{atala_2013, jotzu_2014, aidelsburger_2015, Tan_2018}, Thouless charge pumps \cite{Shuta_2016, Lohse_2016, lohse_2018},  various symmetry-protected topological (SPT) phases \cite{Leder_2016, meier_2016, Bo2018, Xie_2019, leuc_2019, Cai_2019} and quantum spin liquids \cite{semeghini2021}. However, analog quantum simulators are limited by their native interactions to specific models \cite{Altman_2021}.


 Digital quantum simulation offers an approach to realize exotic phases  because in principle, a much wider range of interactions can be synthesized efficiently using the appropriate quantum logic gates \cite{Kitaev_1997}. Already, the preparation of various topological phases has been reported. For example, SPT phases in a spin-1/2 chain model with three-body interactions have been realized on superconducting quantum processors \cite{Choo_2018, Azses_2020, smith2020_crossing}. Similarly, quantum circuits to prepare the ground states of the toric code \cite{satzinger2021} and topological Floquet phases \cite{Mei_2020} have also been developed and used to probe the topological properties of such systems. It is therefore natural to consider using digital quantum computers to study models with interactions beyond nearest-neighbor (NN) interactions as such couplings are reported to host exotic topological phases \cite{Varney2011, Zhu2013} but are difficult to engineer in conventional analog quantum simulators. Early efforts in this direction include simulation of an extended Kitaev chain model with next-nearest neighbor (NNN) interactions on a superconducting quantum device \cite{koh2021}. Another relevant model is a spin-1/2 chain with NNN interactions, which is predicted to host two types of SPT phases with distinct string-order parameters and edge excitations \cite{Zou_2019}. Despite the proposal to prepare these phases using polar molecules, thus far these phases have not been realized on either analog or digital quantum simulators.


Here, we report the  experimental realization of SPT phases of a spin-1/2 chain with NNN interactions on up to 11 qubits of a programmable superconducting quantum processor, Google's Rainbow processor. The model hosts two SPT phases that are distinguished by finite or zero values of different string order parameters. We demonstrate that these phases can be prepared and distinguished by these non-local observables on quantum hardware. Further, we observe edge excitations localized to different sides of the chain depending on the particular phase, confirming another prediction for this system. Our work demonstrates the capabilities of present quantum devices to prepare novel phases with beyond  NN interactions and paves the way to explore other  topological phases on near-term quantum hardware.

\section{Results}

\subsection{SPT phases of spin-1/2 chain with NNN hopping}
\label{sec:model}

We consider a one-dimensional spin-{1/2} chain with NN and NNN interactions as shown in Fig.~\ref{fig:schematic_chain}. Its Hamiltonian is given by

\begin{equation}\label{eqn:SPT}
    H_{T} = -\sum_{k}[J_{1}'(\sigma^{x}_{2k}\sigma^{x}_{2k+1}+\sigma^{y}_{2k}\sigma^{y}_{2k+1})
            +J_{1}(\sigma^{x}_{2k+1}\sigma^{x}_{2k+2}+\sigma^{y}_{2k+1}\sigma^{y}_{2k+2})
            +J_{2}(\sigma^{x}_{k}\sigma^{x}_{k+2} + \sigma^{y}_{k}\sigma^{y}_{k+2})]
\end{equation}
where $\sigma^{x}, \sigma^{y}$ are Pauli operators, $J_{1}'$ ($J_{1}$) denotes the strength of the NN interactions from the even to odd sites (odd to even), and $J_{2}$ denotes the strength of the NNN coupling.

The phase diagram of $H_{T}$ contains two distinct gapped SPT phases known as  the even-parity dimer (ED) and singlet-dimer (SD) phases  \cite{Zou_2019}. The model is in the ED (SD) phase when $J_{1}'=2J_{1}<0,J_{1}>0$ $(J_{1}=-2J_{2}>0,J_{1}'<0)$. These phases can be distinguished by the location of their edge excitations; for a lattice with an odd number of lattice points, the ED (SD) phase has an edge excitation on the right (left) edge of the chain. In addition, the phases can be distinguished by string order parameters, defined as: 

\begin{equation}\label{SOPs}
    O_{n}^{z} = -\textrm{lim}_{r\rightarrow\infty} \langle(\sigma^{z}_{n}+\sigma^{z}_{n+1})e^{i\pi\sum_{k}\sigma^{z}_{k}}(\sigma^{z}_{2r+n}+\sigma^{z}_{2r+n-1}) \rangle \}
\end{equation}

where $\sigma^{z}$ is a Pauli operator, the sum over $k$  is restricted to $n+2\leq k\leq 2r+n-1$ and $r$ is the length of the chain. For this work, the value of $r$ is restricted to a finite value equal to the number of qubits used. Generally, a non-zero string-order parameter indicates the presence of hidden long-range order and a topologically non-trivial phase. In the present model, the ED (SD) phase exhibits a finite $O^{z}_{n}$ value for odd (even) $n$.

\subsection{Preparation of SPT phases on a digital quantum processor}

We used circuits based on adiabatic state preparation \cite{Born_1928, Aspuru_2005} (ASP)  to prepare the SPT phases of $H_{T}$ on a superconducting quantum processor. As described further below,  circuit recompilation \cite{Sun_2021}  was used to decrease the depth of these circuits. The system is initialized in the ground state of an initial Hamiltonian $H_I$  and evolved to the ground state of the target Hamiltonian $H_{T}$ over time duration $T$ using a linear interpolation $ H(s) = (1-s)H_{I} + sH_{T}$ where $s \equiv t/T$. In this study,  $H_{T}$ is given in Eqn. \ref{eqn:SPT}. The initial Hamiltonian, $H_{I}$, is given by

\begin{equation}\label{eqn:Initial Hamiltonian}
    H_{I} = - B_{z}\sum_{k}(-1)^{k}\sigma^{z}_{k}
\end{equation}

with $B_{z}$  a uniform external field. For $B_{z}>0$ and an odd number of sites, the ground state of $H_{I}$ is given by $|0101...01010\rangle$ which can be prepared by applying X-rotation single-qubit gates on sites labeled by odd indices.

\begin{figure}[t!]
\centering{\includegraphics[width=0.8\textwidth]{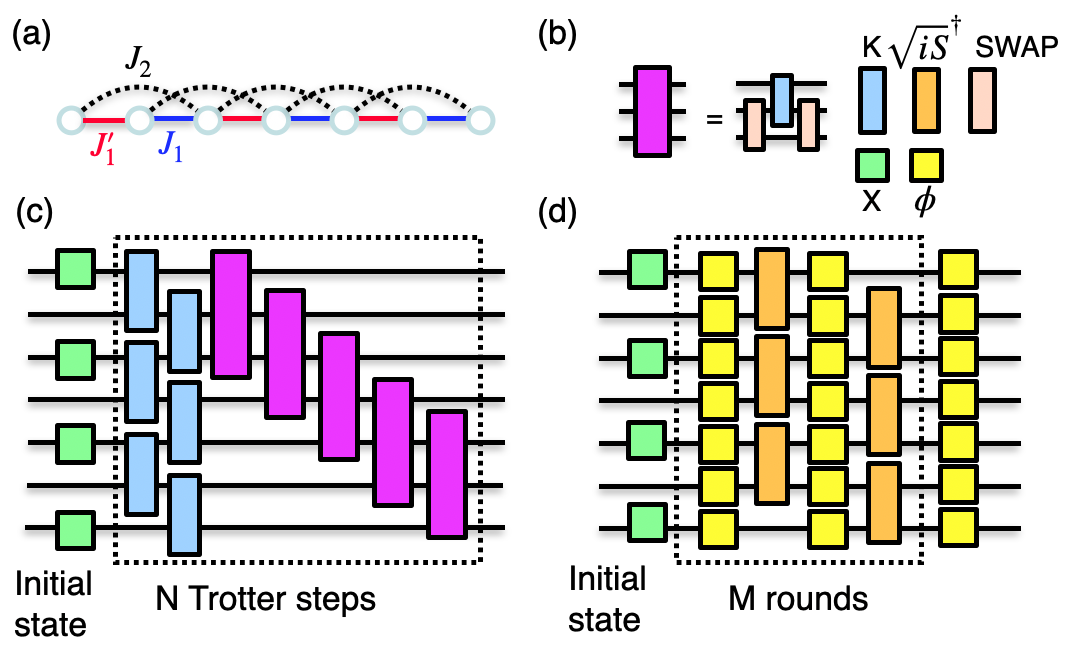}
\phantomsubcaption\label{fig:schematic_chain}
\phantomsubcaption\label{fig:representation}
\phantomsubcaption\label{fig:schematic_trotter}
\phantomsubcaption\label{fig:schematic_recompiled}
}
\caption{\textbf{Schematic of quantum circuits used to prepare the SPT phases.} (\textbf{a}) Arrangement of sites in a 1D-lattice of 7 sites. The strength of interactions going from even-labelled to odd-labelled (odd-labelled to even-labelled) sites is given by $J_{1}'$ ($J_{1}$); those for NNN couplings are given by $J_{2}$. For this study, we also consider chains with 9 and 11 sites. (\textbf{b}) The quantum gates used to construct the circuits for the experiments. Their matrix representations are provided in Ref.~\cite{cirq_developers_2021}. The gates PhasedXZ $(\phi)$ and $\sqrt{iS}^{\dag}$ are the native one-qubit and two-qubit gates on the Google Rainbow processor. (\textbf{c}) Circuit to implement Trotterized ASP for a system with 7 sites. (\textbf{d}) Schematic of the recompiled circuit with $M$ gate rounds.}
\label{fig:schematic}
\end{figure}

To carry out ASP, we Trotterized the adiabatic evolution to first order and implemented the resulting steps using quantum circuits constructed from single-qubit and two-qubit quantum gates, as shown in Fig.~\ref{fig:representation}. The two-qubit gate $K$,  known as the FSIM gate, can be constructed using the native gate set available on the Rainbow processor, Sycamore family \cite{Arute_2020}. An example of a circuit that implements a single Trotter step for a system of 7 sites is shown in Fig.~\ref{fig:schematic_trotter}.

Despite extensive experimentation, the overall circuit that carries out the full ASP was found to be too deep to be accurately implemented. To reduce circuit depth, we used a circuit recompilation scheme \cite{Sun_2021} by fitting the circuits needed to realize the state at each time in the adiabatic evolution to a parameterized circuit. In Ref.~\cite{Sun_2021}, the parameterized circuits consisted of alternating layers of single-qubit gates and two-qubit gates. We used this ansatz in our experiments by using the native gate $\sqrt{\textrm{iS}}^{\dag}$ for the two-qubit gate and the native gate PhasedXZ $(\phi)$ for the single qubit gate, respectively. A schematic of the final recompiled circuit is shown in Fig.~\ref{fig:schematic_recompiled}.


We perform 8192 repetitions of each circuit with measurements in the Z-basis for all sites at each Trotter step. While the quantum circuits implemented conserve the total $z-$component of spin,
$S_z$, the presence of hardware error can lead to $S_z$ non-conservation. We mitigate this error by post-selecting the measurements for $\Delta S_z=0$.

The quantities required to compute string order parameters in Eqn.~\ref{SOPs} can be computed from the measurements in the Z-basis. Similarly, the occupancy of the $j$th site is simply related to the expectation value $\langle \sigma^{z}_{i}\rangle$. Both quantities  can be directly computed by performing the appropriate sums with the measurement bitstrings. Only those bitstrings with $S_z=0$ are used in the computation.

\subsection{Experimental signatures of SPT phases}

We first report calculations of the string order paramters $O_{z_{1}}$ for the ED phase versus ASP time $s$ for spin chains with  7, 9 and 11 sites. We collected data from 15 configurations of qubits; based on the $\sqrt{\textrm{iSWAP}}$ gate cross-entropy benchmarking (XEB) average error per cycle, we selected the ten best  configurations, from which we computed the mean and standard deviation for all observables. To prepare the ED phase, the Hamiltonian parameters were set to $J_{1} = 0.2$, $J_{1}' = -1.5$, $J_{2} = -0.1$, $B_{z} = 2.5$ and $T = 3.0$, and $M=5$ layers of gate rounds were used for circuit recompilation. Emulated results were obtained by running the Trotterized ASP circuit on  Google's circuit emulator qsim \cite{quantum_ai}. 

\begin{figure*}[h!]
\centering
{
\phantomsubcaption\label{fig:7sites}
\phantomsubcaption\label{fig:9sites}
\phantomsubcaption\label{fig:11sites}
\includegraphics[width=0.85\textwidth]{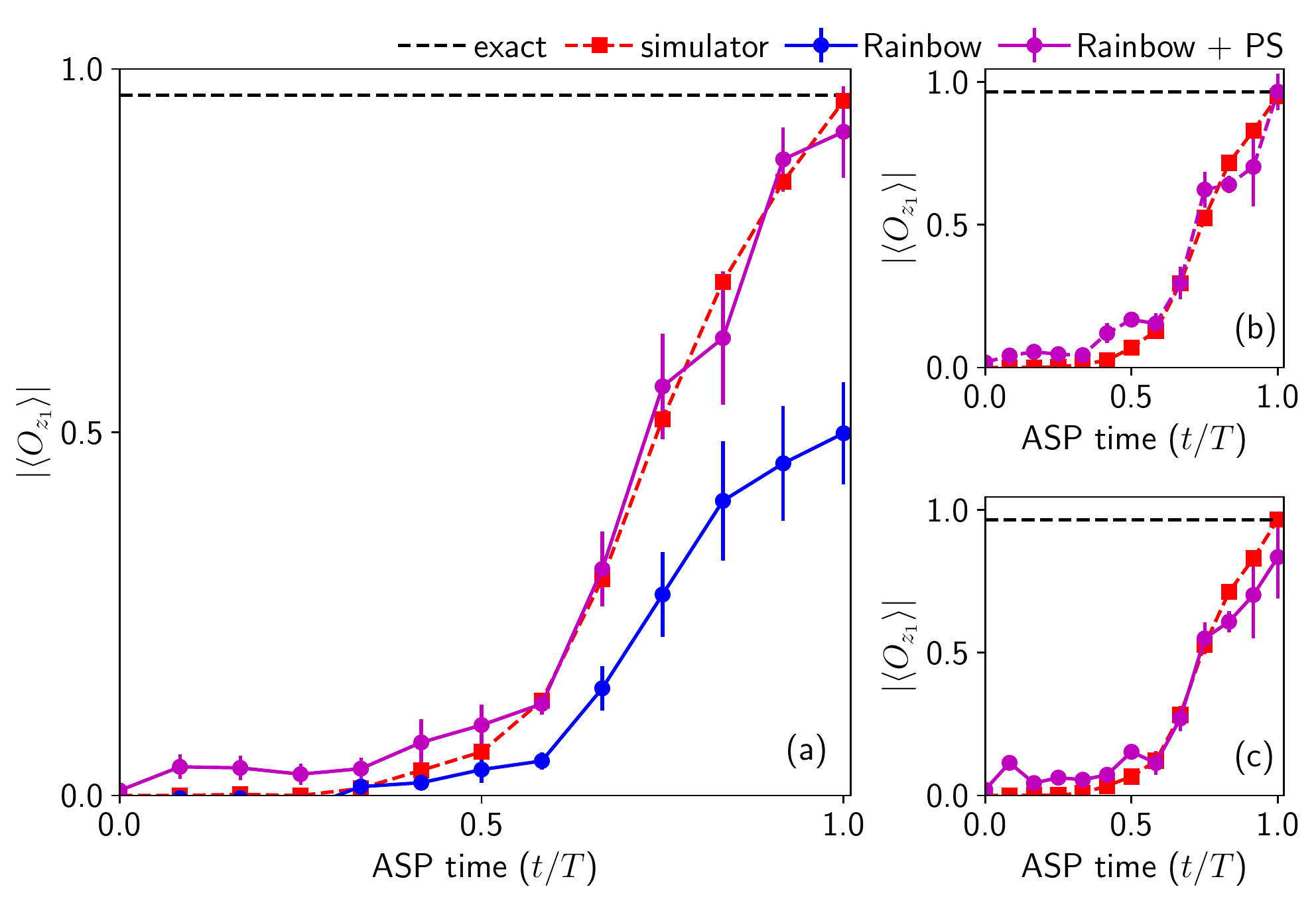}}
\caption{\textbf{Preparation of ED phase for increasing system sizes on Rainbow quantum processor.} Absolute value of the string order parameter $O_{z_{1}}$ versus  ASP time $(s)$ for a system size of (\textbf{a}) 7, (\textbf{b}) 9, and (\textbf{c}) 11 qubits, respectively. Data from Rainbow was collected using 15 different configurations of qubits, and only the best 10 configurations were selected based on their $\sqrt{\textrm{iSWAP}}$ gate XEB average error per cycle. The hardware data without any error mitigation (blue symbols) yields qualitative agreement with the emulated ASP trajectory (red symbols). Quantitative agreement is obtained when post selection is used (purple symbols).  The ED phase can be prepared reliably for system sizes of up to 11 qubits. The parameters $J_{1} = 0.2$, $J_{1}'=-1.5$, $J_{2}=-0.1$, $B_{z}=2.5$, and $T=3.0$ are used to prepare the ED phase. The lines through the symbols are guides to the eye.}
\label{fig:increasing system size}
\end{figure*}

We plot  $|\langle O_{z_{1}}\rangle|$  versus ASP time $s$ on 7 sites in Fig. \ref{fig:7sites}. We observe good agreement between  the final value of  $|\langle O_{z_{1}}\rangle |$ at $s=1$ obtained from Trotterized ASP using qsim and the value from exact diagonalization in Fig.~\ref{fig:7sites}. This result indicates a Trotter step size of $0.25$ is sufficiently small enough to approximate the ASP evolution that yields the ED phase with high fidelity.

We next compare the data obtained by running Trotterized ASP trajectories on qsim with the data obtained by running recompiled circuits on Rainbow without any error mitigation for 7 sites. These circuits required 30 $\sqrt{\textrm{iS}}^{\dag}$ gates. Although the trend of $O_{z_{1}}$ increasing with ASP time is reproduced as seen in Fig.~\ref{fig:7sites},  there is a clear discrepancy in the final value of $O_{z_{1}}$ at the end of the adiabatic trajectory. To mitigate this discrepancy, we perform  post-selection based on $S_z$ conservation. We observe a marked improvement in the quality of the hardware data, with quantitative agreement obtained between the hardware data and the simulator. With this error mitigation step, the quantum processor is able to reproduce the adiabatic trajectory with sufficient fidelity to arrive at the expected non-zero value of the string order parameter  in the ED phase. 

We next compute $O_{z_{1}}$ for system sizes of 9 and 11 qubits. The number of two-qubit gates used in the recompiled circuits was 40 and 50, respectively, compared to 30 in the 7 qubit case. Despite the larger number of gates, we  observe good agreement in the value of the string order parameter over the adiabatic trajectory in Fig. \ref{fig:9sites} and \ref{fig:11sites}, although with a slight degradation that likely arises from the deeper circuits. The data indicates that the SPT phases for a system of 11 sites can be prepared with enough fidelity to observe its topological features on the Rainbow quantum processor.

\begin{figure*}[h!]
\centering
{
\phantomsubcaption\label{fig:ED_Oz0}
\phantomsubcaption\label{fig:ED_Oz1}
\phantomsubcaption\label{fig:SD_Oz0}
\phantomsubcaption\label{fig:SD_Oz1}
\phantomsubcaption\label{fig:ED_site}
\phantomsubcaption\label{fig:SD_site}
\includegraphics[width=\textwidth]{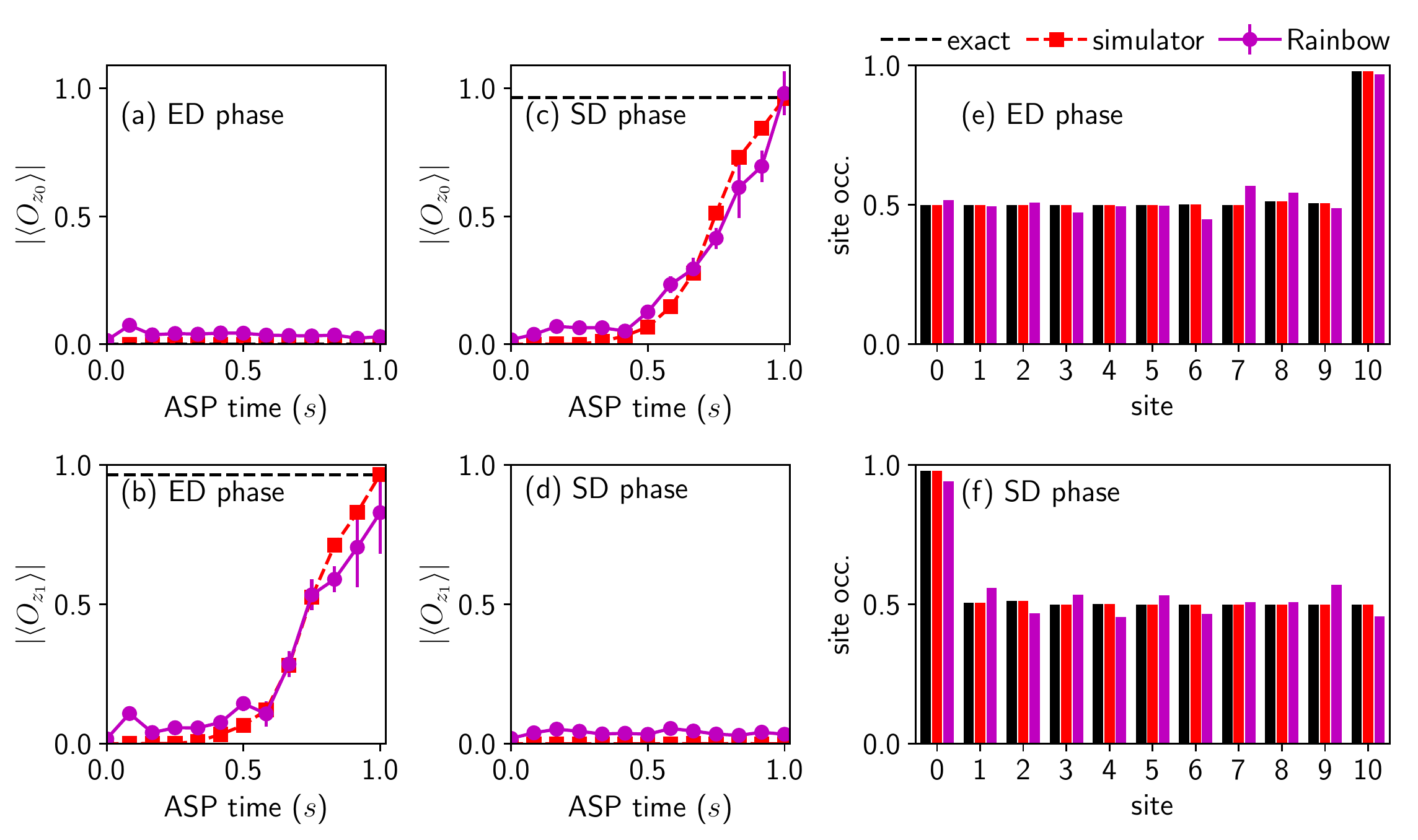}}
\caption{\textbf{Preparation of ED and SD phases using 11 qubits on the Rainbow quantum processor.}  Absolute value of the string order parameters (\textbf{a}) $O_{z_{0}}$ and (\textbf{b}) $O_{z_{1}}$ versus ASP time $(s)$ in the ED phase. (\textbf{c, d}) Analogous result for the SD phase.  Occupancy of each site at the end of the ASP trajectory for the (\textbf{e}) ED and (\textbf{f}) SD phases. The two SPT phases can be prepared and distinguished clearly by finite or zero string-order parameter and location of edge excitation. The parameters $J_{1} = 0.2$, $J_{1}'=-1.5$, $J_{2}=-0.1$, $B_{z}=2.5$, $T=3.0$ were used to prepare the ED phase. The parameters $J_{1} = 1.5$, $J_{1}'=-0.2$, $J_{2}=-0.1$, $B_{z}=2.5$, $T=3.0$ were used to prepare the SD phase. }
\label{fig:SD and ED phases}
\end{figure*}

Next, we verify that we can distinguish the SD and ED phases using the string order parameters. Figure \ref{fig:ED_Oz0} and ~\ref{fig:ED_Oz1} shows  $|\langle O_{z_{0}}\rangle|$ and $|\langle O_{z_{1}}\rangle|$ versus $s$ on 11 qubits  when the model is tuned into the ED phase. We observe good agreement between the hardware data and the simulator over the adiabatic path.  At the end of the adiabatic path, we measure $0.029\pm 0.007$ and $0.829 \pm 0.147$ for $O_{z_{0}}$ and $O_{z_{1}}$, respectively, which is in good agreement with the expected values of $\sim 0$ and $0.964$. Similarly, we tune the model into the SD phase by setting the Hamiltonian parameters to $J_{1} = 1.5$, $J_{1}'=-0.2$, $J_{2}=-0.1$. The string order parameters $O_{z_{0}}$ and $O_{z_{1}}$ versus $s$ are given in  Figs.~\ref{fig:SD_Oz0} and ~\ref{fig:SD_Oz1}, respectively. Again, the final values of the string order parameter from the hardware are $0.981 \pm 0.085$ and $0.034 \pm 0.013$, which are in quantitative agreement with the numerically determined exact values of $0.962$ and $\sim 0$. In both cases, we measured a finite value for the appropriate string order parameters and nearly zero for the other, indicating that the correct SPT phases were successfully prepared.  

Finally, we plot the occupancy of each site at the end of the adiabatic evolution for the ED and SD phases in Figs.~\ref{fig:ED_site} and \ref{fig:SD_site}, respectively. In the ED (SD) phase, an edge excitation is predicted to exist on the right (left) end of the chain. This feature is indeed observed using the exact solution obtained from exact diagonalization. The results from the hardware clearly indicate a difference in the occupancy on the appropriate edge of the chain for each phase and the rest of the chain, with the value in good agreement with the exact result. This observation provides additional evidence that the SPT states prepared on the hardware exhibit the key features expected of these topological phases.


\section{Discussion}

We have reported the preparation of SPT phases of a spin-1/2 chain with NNN interactions on up to 11 qubits. We observed edge excitations localized to a specific end of the chain depending on the specific phase and obtained good agreement between the measured string order parameters $O_{z_{0}}$ and $O_{z_{1}}$ and the numerically exact values. These observations indicate that topological phases of spin models with beyond NN interactions can be prepared on near-term devices using ASP with good fidelity. We note that this work required the use of  circuit recompilation techniques to reduce the gate depth for ASP. Without circuit recompilation, the number of required two-qubit gates was around 170 for 7 sites, yielding only qualitative agreement with the expected final string order parameter value.  Numerical experimentation  indicates that the origin of this discrepancy is partly due a parasitic controlled phase in the native two-qubit $\sqrt{\textrm{iS}}^{\dag}$ \cite{Arute_2020} that results in unwanted interactions in the adiabatic trajectory (see Supplementary Information Section I). Efforts to mitigate this phase were unsuccessful (see Supplementary Information Section II).


Future work may therefore focus on mitigating parasitic interactions along with further increasing the fidelity of the two-qubit gates so as to enable circuits with hundreds of two-qubit gates. Although the execution of such circuits have been reported for some tasks \cite{Arute_2020}, we find that the maximum number of two-qubit gates that can be used while retaining quantitative agreement of observables depends on the specific problem under consideration. With such advances, we anticipate that the preparation of other exotic topological phases may be feasible, such as spin-{1/2} systems in two dimensions with similar couplings which are thought to host chiral spin liquids.
\cite{Varney2011, Zhu2013, Sedrakyan2015}.






\section{Materials and Methods} \label{Materials and Methods}
We used the Google Rainbow processor, Sycamore family for this study. The processor consists of a two-dimensional array of 23 transmon qubits with each qubit  tunably coupled to its neighbors. The native single-qubit gates are the PhasedXZ gate which consists of a rotation about an axis in the XY plane of the Bloch sphere with an extra phase about the Z axis. The native two-qubit gates are the $\sqrt{\textrm{iS}}^{\dag}$ gates. Further details of the processor are available in Ref.~\cite{Arute_2020}.

\label{sec:methods}

\section*{Acknowledgements}
The authors thank M. Wojciech, N.~C. Rubin, J. Zhang and  R. Babbush for helpful discussions. \textbf{Funding:} A.T.K.T, S.-N.S., A.J.M and G.K.-L.C were supported by the US NSF under Award No.~1839204. R.~N.~Tazhigulov was supported by the US Department of Energy, Office of Basic Energy Sciences, under Award No.~DE-SC0019374. \textbf{Author contributions:} A.T.K.T and A.J.M. conceptualized the project. A.T.K.T designed and optimized the circuit and executed simulations with assistance from S.S. and R.N.T. A.T.K.T carried out the simulation runs on the quantum processor and analyzed the results. A.T.K.T and A.J.M. wrote the paper. All authors discussed the results and contributed to the development of the manuscript. \textbf{Competing interests:} The authors declare that they have no competing interests.
\newpage
\bibliography{main}

 \newpage
 \setcounter{section}{0}
 \setcounter{equation}{0}
 \setcounter{figure}{0}
 
\renewcommand{\figurename}{Fig.}
\renewcommand{\thefigure}{S\arabic{figure}}

\renewcommand{\theequation}{S.\arabic{equation}}
 
 \title{\textmd{Supplementary materials for}\\Realizing symmetry-protected topological phases in a spin-1/2 chain with next-nearest neighbor hopping on superconducting qubits}
\setcounter{page}{1}

\begin{minipage}[h]{\textwidth}
    \maketitle
\end{minipage}

\noindent \textbf{This PDF file includes:} \\
\indent Sections SI to SII \\
\indent Figs. SI to SII
\vfill
\small{*aminnich@caltech.edu}
\newpage
\centerline{\textbf{Supplementary materials for}}
\section{Role of parasitic controlled phase}

\begin{figure*}[h!]
\centering
{
\phantomsubcaption\label{fig:sweep_phi}
\phantomsubcaption\label{fig:sweep_gamma}
\phantomsubcaption\label{fig:sweep_zeta}
\phantomsubcaption\label{fig:sweep_chi}
\includegraphics{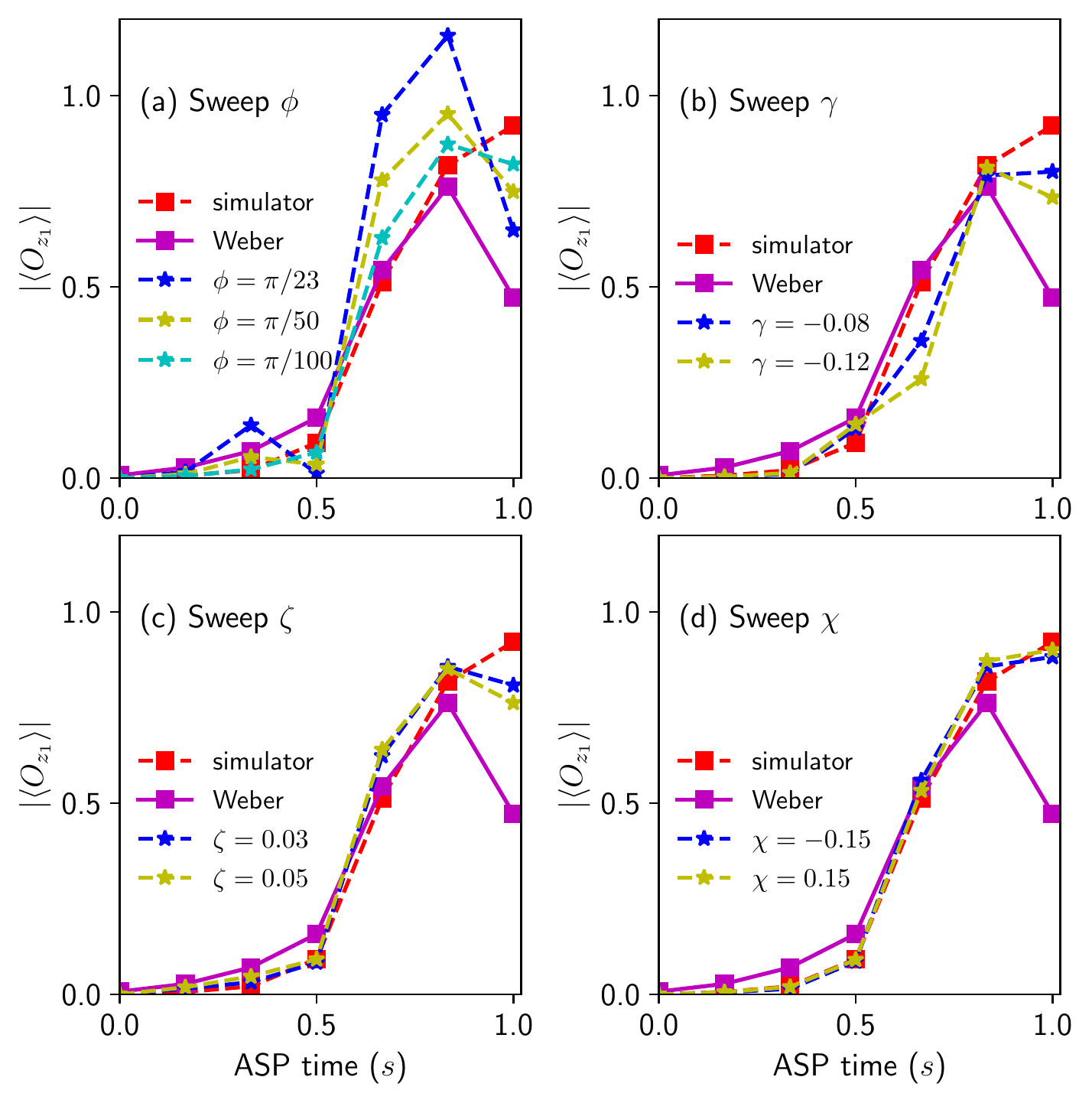}}
\caption{\textbf{Effects of gate imperfections on $|\langle O_{z_{1}}\rangle|$ in the ED phase.}  Absolute value of the string order parameters $O_{z_{1}}$ versus ASP time $(s)$ in the ED phase when (\textbf{a}) $\phi$, (\textbf{b}) $\gamma$, (\textbf{c}) $\zeta$, and (\textbf{d}) $\chi$ is varied. As reference, we also plot the data obtained from Weber and the noiseless data from the simulator. The string order parameter is most sensitive to  $\phi$. The parameters $J_{1} = 0.2$, $J_{1}'=-1$, $J_{2}=-0.1$, $B_{z}=1.5$, $T=3.0$ were used to prepare the ED phase. }
\label{fig:numerical_simulation}
\end{figure*}

This section presents a numerical investigation of the effects of systematic errors in the two-qubit gate on non-recompiled circuits that implement digitized adiabatic preparation of the SPT phase. The most general excitation-number-conserving two-qubit gate takes the following form (with the basis states in the order $|00\rangle$, $|01\rangle$, $|10\rangle$, and $|11\rangle$):

\begin{equation}
    U(\theta,\zeta ,\chi,\gamma,\phi) =
    \begin{pmatrix}
    &1 &0 &0 &0\\
    &0 &e^{-i(\gamma+\zeta)}\textrm{cos}\theta &-ie^{-i(\gamma-\chi)}\textrm{sin}\theta &0\\
    &0 &ie^{-i(\gamma+\chi)}\textrm{sin}\theta &e^{-i(\gamma-\zeta)}\textrm{cos}\theta &0\\
    &0 &0 &0 &e^{-i(2\gamma + \phi)}
    \end{pmatrix}
\end{equation}
While the ideal native two-qubit gate on Rainbow and Weber is given by $U(\pi/4,0,0,0,0)^{\dag}$, additional interactions lead to non-zero values of $\zeta$, $\chi$, $\gamma$ and $\phi$. We numerically simulate the effects of these non-idealities on the value of the string order parameter along the adiabatic trajectory by plotting $|\langle O_{z_{1}}\rangle|$ against ASP time $s$ on 7 sites for different values of $\phi$, $\gamma$, $\zeta$, and $\chi$.   Figure \ref{fig:numerical_simulation} shows the results for representative values of  $\phi$, $\gamma$, and $\zeta$. We also plot results obtained from Weber and the ideal trajectory obtained using emulations. We observe the biggest variation in the string order parameter when $\phi$ is varied. Further, the qualitative trend of non-monotonic variation of the string order parameter towards the end of the adiabatic trajectory agrees with that observed experimentally. These observations suggest that the parasitic controlled phase $\phi$ is the origin of the discrepancies for circuits of sufficient depth.

\clearpage
\newpage
 \section{Attempts to mitigate the effects of parasitic controlled phase in non-recompiled circuits}
 
This section describes  attempted strategies to mitigate the parasitic controlled phase. 

 \subsection{Constructing an exact CPHASE gate using two noisy native two-qubit gates}
 
 The first approach constructs the $\textrm{CPHASE}(-\phi)$ gate exactly using a series of single-qubit rotations and two $\sqrt{iS}^{\dag}_{hardware}$ to compensate for the phase in each $\sqrt{iS}^{\dag}_{hardware}$. The cost of this approach is the addition of two native two-qubit gates for each original two-qubit gate, thereby increasing the gate depth by a factor of 3.

 
 We tested this scheme on Weber by performing Floquet characterization to estimate the $\phi$ present on each qubit \cite{Arute_2020} and used the average value to construct a compensated $\textrm{CPHASE}(-\phi_{avg})$. The results of the compensated circuits is presented in Fig.~\ref{fig:accounting_for_phi}. We observe  greater deviations from the exact result when the compensated circuits are used. The likely origin of the worse  performance is the  larger number of two-qubit gates used in the compensated circuits (510 versus 170 to reach the end of the adiabatic path).
 
 
 \subsection{Single qubit Z-rotations}
\begin{figure*}[h!]
\centering
{
\phantomsubcaption\label{fig:accounting_for_phi}
\phantomsubcaption\label{fig:single_rotations}
\includegraphics{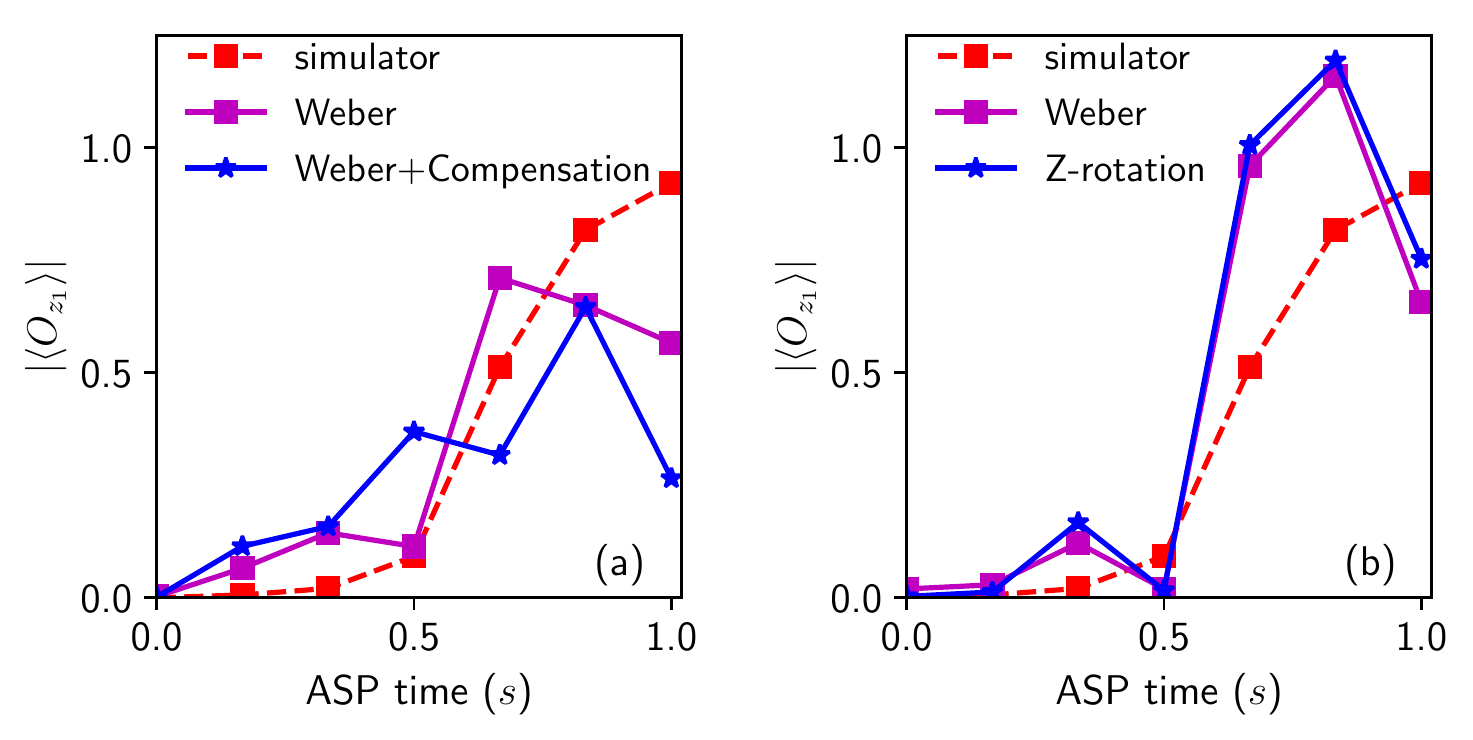}}
\caption{\textbf{Attempts to compensate for parasitic controlled phase.} Absolute value of the string order parameters $O_{z_{1}}$ versus ASP time $(s)$ in the ED phase for circuits with (\textbf{a}) $\textrm{CPHASE}(\phi)$ appended to each native two-qubit gate to compensate for the parasitic controlled phase; (\textbf{b}) single-qubit Z rotations are added to split the parasitic phase among two basis states. The data obtained from Weber and the noiseless data from the simulator are also shown. The qualitative trend of the string order parameter is mostly unchanged by the two strategies. The parameters $J_{1} = 0.2$, $J_{1}'=-1$, $J_{2}=-0.1$, $B_{z}=1.5$, $T=3.0$ were used to prepare the ED phase.}
\label{fig:comepnsation}
\end{figure*}

 The second approach is based on the observation that the phase present in the $|11\rangle$ state can be removed at the expense of adding half the phase to the $|01\rangle$ state and $|10\rangle$ state. Specifically, single-qubit Z-rotations with an angle of $-\phi/2$ are applied to each qubit before applying the native two-qubit gate.  Given that fidelity is a quadratic function of gate parameters, a higher fidelity can be obtained by splitting the phase into two. We tested this scheme by performing Floquet calibration to estimate the $\phi$ present on each qubit and used the average to perform single-qubit Z rotations on the qubits. The result is shown in Fig.~\ref{fig:single_rotations}. Although some improvement in the final value of the string order parameter is observed, the non-monotonic trend remains largely unchanged, indicating that manipulation of the parasitic phase is inadequate to remove the discrepancy.

 
\end{document}